# Time Series Management Systems: A Survey

Søren Kejser Jensen, Torben Bach Pedersen, *Senior Member, IEEE*, Christian Thomsen

**Abstract**—The collection of time series data increases as more monitoring and automation are being deployed. These deployments range in scale from an Internet of things (IoT) device located in a household to enormous distributed Cyber-Physical Systems (CPSs) producing large volumes of data at high velocity. To store and analyze these vast amounts of data, specialized *Time Series Management Systems (TSMSs)* have been developed to overcome the limitations of general purpose Database Management Systems (DBMSs) for times series management. In this paper, we present a thorough analysis and classification of TSMSs developed through academic or industrial research and documented through publications. Our classification is organized into categories based on the architectures observed during our analysis. In addition, we provide an overview of each system with a focus on the motivational use case that drove the development of the system, the functionality for storage and querying of time series a system implements, the components the system is composed of, and the capabilities of each system with regard to *Stream Processing* and *Approximate Query Processing (AQP)*. Last, we provide a summary of research directions proposed by other researchers in the field and present our vision for a next generation TSMS.

## 1 INTRODUCTION

THE increase in deployment of sensors for monitoring large industrial systems and the ability to analyze the collected data efficiently provide the means for automation and remote management to be utilized at an unprecedented scale [1]. For example, the sensors on a Boeing 787 produce upwards of half a terabyte of data per flight [2]. While the use of sensor networks can range from an individual smart light bulb to hundreds of wind turbines distributed throughout a large area, the readings from any sensor network can be represented as a sequence of values over time, more precisely as a *time series*. Time series are finite or unbounded sequences of data points in increasing order by time. *Data series* generalize the concept of time series by removing the requirement that the ordering is based on time. As time series can be used to represent readings from sensors in general, the development of methods and systems for efficient transfer, storage, and analysis of time series is a necessity to enable the continued increase in the scale of sensor network and their deployment in additional domains [1], [3], [4], [5]. For a general introduction to storage and analysis of time series see [6], [7], a more in-depth introduction to sensor data management, data mining, and stream processing is provided by [8], [9], [10].

While general Database Management Systems (DBMSs), and in particular Relational Database Management Systems (RDBMSs), have been successfully deployed in many situations, they are unsuitable to handle the velocity and volume of the time series produced by the large scale sensor networks deployed today [3], [4], [5], [11]. In addition, analysis of the collected time series often requires exporting the data to another application such as R or SPSS, as these provide additional capabilities and a simpler interface for time series analysis compared to an RDBMS, adding complexity to the analysis pipeline [12]. In correspondence with the increasing need for systems that efficiently store and analyze time series, *Time Series Management Systems (TSMSs)*[1] have been proposed for multiple domains including monitoring of industrial machinery, analysis of time series collected from scientific experiments, embedded storage for Internet of things (IoT) devices, and more. For this paper we define a TSMS as any system developed or extended for storing and querying data in the form of time series. Research into TSMSs is not a recent phenomenon and the problems using RDBMSs for time series have been demonstrated in the past. In the 1990s Seshadri et al. developed the system SEQ and the SQL-like query language SEQUIN [13]. SEQ was built specifically to manage sequential data using a data model [14] and a query optimizer that utilize that the data is stored as a sequence and not a set of tuples [15]. SEQ was implemented as an extension to the object-relational DBMS PREDATOR with the resulting system supporting storage and querying of relational and sequential data together. While additional support for sequences was added to the SQL standard through for examples window queries, development of query languages and TSMSs continued throughout the early 2000s. Lerner et al. proposed the algebra and query language AQuery [16] for which data is represented as sequences that can be nested to represent structures similar to tables. Utilizing the AQuery data model and information provided as part of the query, such as sort order, novel methods for query optimization were implemented with improved query performance as a result. In contrast to AQuery which uses a non-relational data model, Sadri et al. proposed an extension to the SQL query language named SQL-TS for querying sequences in the form of time series stored in an RDBMS or from an incoming stream [17]. SQL-TS extended SQL with functionality for specifying which column uniquely identifies a sequence and which column the sequence should be ordered by. Patterns to search for in each sequence can then be expressed as predicates in a WHERE clause. They proposed a query optimizer named OPS based on an existing string search algorithm, making complex pattern matching both simple to express and efficient to execute [18], [19].

---

• *S. K. Jensen, T. B. Pedersen, and C. Thomsen, are with the Department of Computer Science at Aalborg University, Denmark. E-mail: {skj, tbp, chr}@cs.aau.dk.*

---

1. TSMS is one among multiple names for these systems commonly used in the literature, another common name is Time Series Database.

It is clear that decades of research into management of time series data have lead to the development of expressive query languages and efficient query processing engines. However, this past generation of TSMSs are unable to process the amount of time series data produced today, as support for parallel query processing is limited and the capability to utilize distributed computing is non-existing. In addition, as the TSMSs developed were designed for general-purpose use, only limited optimizations could be implemented for specific use cases [11]. As a result, as new use cases and technologies appear, such as data management for IoT devices and commodity distributed computing, a new generation of TSMSs have been developed. With this paper we provide an overview of the current research state-of-the-art in the area of TSMSs presented as a literature survey with a focus on the contributions of each system. The goal of this survey is to analyze the current state-of-the-art TSMSs, discuss the limitations of these systems, analyze research directions proposed by other researchers in the field, and as a result of our analyses present our vision for a next generation TSMS. To focus the survey, we primarily analyze systems designed for storing numerous time series persistently, allowing the system to support *aggregate analysis* of multiple data streams over time. In addition, while *scalability* has not been used as a criterion for excluding systems, we see it as an important characteristic for any TSMS and a focus of this survey. As a consequence of these decisions, systems such as Antelope [20], HybridStore [21], LittleD [22] and StreamOp [23] developed for deployment on sensor nodes, and more broadly research in the area of query processing inside sensor networks, are not included. For an introduction to the topic of query processing in sensor networks see [8]. Also, the survey only includes systems with papers published in the *current decade*, due to the switch in focus towards Big Data systems built using commodity hardware for large scale data storage and analytics. For TSMSs we observed a general trend towards *distributed storage and processing*, except for TSMSs developed for evaluation of research or for use in embedded systems. Furthermore, only systems implementing methods specifically for storage and querying of time series are included due to the effect design decisions for each component have on the other. The survey also provides an overview of open research questions in order to provide not just an overview of existing systems but also provide insight into the next generation of TSMSs. Last, in addition to the systems proposed by academia or produced through industrial research, many open-source and proprietary TSMSs have been developed, with the systems OpenTSDB, KairosDB, Druid, InfluxDB, and IBM Informix being popular examples. For information about selecting a TSMS for a particular workload see the following papers [12], [24], [25], [26], [27], [28], [29].

The search method for the survey is a follows. An unstructured search using Google Scholar was performed to provide an overview of what research is performed in the area of TSMSs, and to determine the relevant conferences, terms specific to the research area, and relevant researchers. Based on the papers found during the unstructured search, iterations of structured search were performed. Relevant publications found in each iteration were used as input for the next iteration. In each iteration additional papers were located by exhaustively going through the following sources for each paper:

- The references included in the paper.
- Newer publications citing the paper, the citations were found using Google Scholar.
- All conference proceedings or journal issues published in this decade for the conference or journal in which the paper was presented or published. For the data management outlets the most commonly used were ACM SIGMOD, IEEE Big Data and PVLDB, while papers were primarily presented at USENIX conferences when looking at system outlets.
- The publication history for each paper's author found using a combination of DBLP, Google Scholar and profile pages hosted by the author's employer.

The rest of the paper is organized as follows. Section 2 provides a summary of all the systems included in this survey in addition to a description of the criteria each system will be classified by. Section 3, 4, and 5, describe the systems and are organized based on if the data store is internal, external, or if the system extends an RDBMS. Section 6 provides an overview of open research questions proposed by leading researchers in the field in addition to how these ideas correspond to the surveyed systems, culminating in our vision for a next generation system for time series management and what research must be performed to realize it. Last, a summary and conclusion are presented in Section 7.

## 2 CLASSIFICATION CRITERIA

An overview of the TSMSs included in this survey is shown with regards to our classification criteria in Table 1 and the operations each system supports in Table 2. As some publications refrain from naming the proposed system some systems are marked as *unnamed* with only the references as identification. The TSMSs were organized into three categories based on how the data processing and storage components are connected, due to the major impact this architectural decision has on the implementation of the system. In addition to the differences in architecture, the remaining classification criteria were selected based on what differentiated the surveyed systems from each other. For systems incorporating an existing DBMS only additional functionality described in the surveyed papers will be documented. The full set of classification criteria are:

**Architecture:** The overall architecture of the implementation is primarily determined by how the data store and data processing components are connected. For some systems both the data store and processing engine are *internal* and integrated together in the same application, either due to them being developed together or if an existing DBMS is embedded and accessed through the provided interface. Other systems use an existing *external* DBMS or Distributed

TABLE 1
Summary of the Surveyed Systems in Terms of Classification Criteria

| System | Year | Purpose | Motivational Use Case | Maturity | Distributed | Scale Shown | Processing Engine | API | Approximation | Stream Processing | Storage Engine | Storage Layout |
|---|---|---|---|---|---|---|---|---|---|---|---|---|
| **Internal Stores** | | | | | | | | | | | | |
| SciDB [30] | 2012 | Monitoring | Monitoring of large computer networks with high resolutions. | Mature System | Centralized | 6.3 GB / 1 Node | Implemented (C) | Client Library (C) | No support for approximation of time series. | No-stream processing. | BerkeleyDB | Fixed sized arrays stored as values in BerkeleyDB. |
| FAQ [31] | 2014 | Evaluation | Efficient approximate queries on time series of histograms. | Proof-of-Concept Implementation | Centralized | 19 GB / 1 Node | Implemented (Java) | Unknown | Model-based AQP. | No-stream processing. | KyotoCabinet | Sketches and histograms organized in a range tree. |
| WearDrive [32],[33] | 2015 | IoT | Power efficient storage of data produced by wearable sensors. | Demonstration System | Distributed | Unknown / 2 Nodes | Implemented (C, Java, JNI) | Client Library (C, Java, JNI) | No support for approximation of time series. | Register callbacks waiting for sensor readings. | Implemented (C, Java, JNI) | In-memory log of KV-pairs indexed by a hash table. |
| RINSE [34],[35],[36] | 2015 | Data Analytics | Querying time series without constructing a full index first. | Demonstration System | Centralized | 1 TB / 1 Node | Implemented (C) | Drawn nearest neighbor search. | No support for approximation of time series. | No-stream processing. | Implemented (C) | ADS+ tree indexing an unspecified ASCII format. |
| Unnamed [57] | 2015 | Evaluation | Fast approximated queries for decision support systems. | Proof-of-Concept Implementation | Centralized | Unknown / 1 Node | Implemented (R) | Extended SQL | Model-based AQP used for aggregate queries. | No-stream processing. | Implemented (R) | Separated storage of both raw data and models. |
| Plato [38] | 2015 | Data Analytics | Simple analysis of spatiotemporal data with signal processing. | Demonstration System | Centralized | Unknown / 1 Node | Implemented (Unknown) | Extended SQL | Model-based AQP. | No-stream processing. | Implemented (Unknown) | Tables with models stored as binary large objects. |
| Chronos [39],[40] | 2016 | Monitoring | Monitor hydroelectric plants using PCs with flash memory. | Demonstration System | Centralized | 11.5 GB / 1 Node | Implemented (C++) | Client Library (C++) | No support for approximation of time series. | Out-of-order inserts. | Implemented (C++) | B-Tree index over quasi-sequential data points. |
| Pytsms [41],[42] | 2016 | Evaluation | Reference implementation of two formalisms for time series. | Proof-of-Concept Implementation | Centralized | Unknown / 1 Node | Implemented (Python) | Client Library (Python) | User-defined aggregates at multiple resolutions. | No-stream processing. | Implemented (Python) | Python objects pickled to a file or serialized to CSV. |
| PhilDB [43] | 2016 | Data Analytics | Storage and analysis of versioned mutable time series. | Demonstration System | Centralized | 119.12 MB / 1 Node | Implemented (Python, Pandas) | Client Library (Python) | No support for approximation of time series. | No-stream processing. | Implemented (Python, SQLite) | Triples stored as binary files, updates as HDF5. |
| **External Stores** | | | | | | | | | | | | |
| TSDS [44] | 2010 | Data Analytics | Querying data in multiple formats from one end-point. | Mature System | Centralized | Unknown / 1 Node | Implemented (Java) | REST API serving multiple formats. | AQP using sampling. | No-stream processing. | Implemented (Java) for caching. | A binary file with 64-Bit values, metadata as NcML. |
| SciDB [45],[46],[47] | 2013 | Data Analytics | Storage and querying of data from scientific instruments. | Mature System | Distributed | Unknown / 16 Nodes | Implemented (C++), SciLAPACK | Array languages AFL and AQP. | AQP using sampling. | Arrays piped through external processes. | Implemented (C++), PostgreSQL | N-dimensional arrays of tuples, stored in chunks. |
| Respawn [48],[49] | 2013 | Monitoring | Executing low latency queries across cloud and sensor nodes. | Demonstration System | Distributed | 2 GB / 10,000 Nodes | Bodytrack Datastore | REST API serving JSON. | Predefined aggregates at multiple resolutions. | No-stream processing. | Bodytrack Datastore | Bodytrack Datastore's compressed binary format. |
| SensorGrid [50] | 2013 | Data Analytics | OLAP of sensor data through AQP and grid computing. | Demonstration System | Distributed | Unknown / 5 Nodes | Implemented (Unknown) | Web interface, SQL. | User-defined aggregates at multiple resolutions. | Window queries. | RDBMS | Data points, aggregates in two-dimensional array. |
| Unnamed [51],[52],[53] | 2014 | Evaluation | Indexing mathematical models in a distributed KV-store. | Proof-of-Concept Implementation | Distributed | 12 GB / 9 Nodes | Apache Hadoop | Unknown | Model-based AQP. | Real-time modelling of segmented time series. | Apache HBase | In-memory binary trees indexing models in HBase. |
| Tristan [54],[55] | 2014 | Data Analytics | Enable creation of analytical applications using sensor data. | Demonstration System | Centralized | Unknown / 1 Node | HYRISE | Unknown | Model-based AQP. | Real-time modelling of segmented time series. | HYRISE | Sparse representation after dictionary compression. |
| Druid [56] | 2014 | Monitoring | Ingestion and exploration of metrics in real-time. | Mature System | Distributed | 37-89 MB / 6 Nodes | Implemented (Java) | REST API serving JSON. | User-defined aggregates and model-based AQP. | No-stream processing. | Implemented (Java), DFS | Immutable columns are compressed based on their type. |
| Unnamed [57] | 2014 | Monitoring | Query both time series and relational data through SQL. | Mature System | Distributed | 0.90 TB / 1 Node | IBM Informix | REST API serving JSON. | Model-based AQP. | Real-time modelling of segmented time series. | IBM Informix | Blobs of values, or timestamp deltas and values. |
| Unnamed [58] | 2014 | Monitoring | Remote monitoring of industrial installations in real-time. | Mature System | Distributed | 5 TB / 46 Nodes | GE Streaming Engine, Pivotal GemFire | OQL and Client Library (Java) | No support for approximation of time series. | Real-time data transformations and analytics. | Pivotal GemFire | Ordered KV-pairs storing segments as linked lists. |
| Bolt [59] | 2014 | IoT | Simplify management of IoT data from connected homes. | Demonstration System | Distributed | Unknown / 16 Nodes | Implemented (C#) | Client Library (C#) | AQP using sampling. | Data is written to and read from streams. | Implemented (C#), S3, Azure | In-memory index of continuous chunks on disk. |
| Storacle [60],[61] | 2015 | Monitoring | Enable local processing of data at the edge of sensor networks. | Demonstration System | Distributed | 541 MB / 1 Node | Implemented (Java), Cloud | Unknown | Predefined aggregates. | Online computation of simple aggregates. | Implemented (Java), Cloud | Unknown for flat-format and Protocol Buffers. |
| Gorilla [62] | 2015 | Monitoring | Reducing Query latency for a real-time monitoring system. | Mature System | Distributed | 1.3 TB / 20 Nodes | Implemented (C++) | Client Library (Unknown) | No support for approximation of time series. | No-Stream processing. | Implemented (C++), DFS, HBase. | Blocks of deltas prefixed by a single timestamp. |
| Unnamed [63] | 2015 | Data Analytics | Execute aggregate queries on resource constrained systems. | Mature System | Distributed | 3 TB / 1 Node | Implemented (Java), MySQL | SQL | Model-based AQP used for aggregate queries. | Online computation of aggregate models. | MySQL | Binary tree storing aggregates as models or sums. |
| servioticy [64],[65] | 2015 | IoT | Integration of stream processing and data storage for IoT. | Demonstration System | Distributed | Unknown / 16 Nodes | Couchbase, Apache Storm, Elasticsearch | REST API serving JSON. | No support for approximation of time series. | Apache Storm extended with versioning of bolts. | Couchbase | JSON documents with ids indexed by Elasticsearch. |
| RTODB [66],[67] | 2016 | Monitoring | Analyzing data with ms timestamps at multiple resolutions. | Mature System | Distributed | 2.757 TB / 2 Nodes | Implemented (Go) | Client Library (Go, Python) | Predefined aggregates at multiple resolutions. | Data is written to and read from streams. | DFS, MongoDB | Versioned tree with aggregates in internal nodes. |
| **RDBMS Extensions** | | | | | | | | | | | | |
| TimeTravel [68],[69] | 2012 | Data Analytics | Continuos forecasting of power consumption in a smart grid. | Demonstration System | Centralized | Unknown / 1 Node | PostgreSQL | Extended SQL | Model-based AQP. | No-stream processing. | PostgreSQL | Data points in arrays with layers of models on top. |
| F²DB [70],[71] | 2012 | Data Analytics | Forecasting directly integrated as part of a data warehouse. | Demonstration System | Centralized | Unknown / 1 Node | PostgreSQL | Extended SQL | Model-based AQP for forecast queries. | No-stream processing. | PostgreSQL | Tables and a hierarchy of models built on top. |
| Unnamed [72],[73],[74] | 2016 | Evaluation | OLAP analysis of time series as ordered sequences of events. | Proof-of-Concept Implementation | Centralized | 23 MB / 1 Node | Oracle RDBMS | Extended SQL | Model-based AQP for interpolated data. | No-stream processing. | Oracle RDBMS | Tables for both raw data and model parameters. |

File System (DFS) as a separate data store, requiring the TSMS to implement methods for transferring time series data to the external DBMS or DFS. Last, some implement methods for processing time series as *extensions to an existing RDBMS*, making the RDBMS's internal implementation accessible for the TSMS in addition to removing the need to transfer data to an external application for analysis. As Table 1 is split into sections based on architecture, the architecture used by the systems in each section of the table is written in italics as the heading for that section of the table.

**Year:** The year the latest paper documenting the surveyed system was published. This is included to simplify comparison of systems published close to each other. The year of the latest publication is used as new publications indicate that functionality continues to be added to the system.

**Purpose:** The type of workload the TSMS was designed and optimized for. The first type of systems are those designed for collecting, analyzing and reacting to data from *IoT* devices, next TSMSs developed for *monitoring* large industrial installations for example data centers and wind turbines, then systems for extracting knowledge from time series data through *data analytics*, and last systems for which no other real-world use case exist than the *evaluation* of research into new formalisms, architectures, and methods for TSMSs.

**Motivational Use Case:** The intended use case for the system based on the problem that inspired its creation. As multiple systems designed for the same purpose can be designed with different trade-offs, the specific problem a system is designed to solve indicates which trade-offs were necessary for a particular TSMS.

**Distributed:** Indicates if the system is intended for a *centralized* deployment on a single machine or built to scale through *distributed* computing. It is included due to the effect this decision had on the architecture of the system and the constraints a centralized system adds in terms of scalability.

**Maturity:** The maturity of the system based on a three level scale: *proof-of-concept implementations* implement only the functionality necessary to evaluate a new technique, *demonstration systems* include functionality necessary for users to interact with the system and are mature enough for the system to be evaluated with a real-life use case, *mature systems* have implementations robust enough to be deployed to solve real-life use cases and supported through either an open-source community or commercial support.

**Scale Shown:** Scale shown is used as a concrete measure of scalability and defined as the *size of the largest data set* and *number of nodes* a TSMS can manage as documented in the corresponding publications. The sizes of the data sets are reported in bytes after the data has been loaded into the TSMS. As data points differ in size depending on the data types used and the amount of metadata included, data sets for which the size is documented as the number of data points are not included. While some systems are continuously being improved, only the results presented in the referenced publications are included for consistency.

**Processing Engine:** The data processing engine used by the system for querying, and if supported, analyzing the stored time series data. Included as this component provides the system's external interface for the user to access the functionality provided by the system. If an existing system is used we write the *name of the system*, otherwise if a new engine has been developed we write *implemented* with the implementation language added in parentheses.

**API:** The primary methods for interacting with the system. The methods can be *query languages*, *extensions of existing query languages*, *client libraries* developed using general purpose programming languages, a *web service*, or a *web interface*.

**Approximation:** Describes the primary method, if any, for approximating time series that each system supports. Using approximation as part of a TSMS provides multiple benefits. Representing time series as simple aggregates uses less storage and reduces query processing time for queries capable of utilizing the aggregate. Approximate Query Processing (AQP) expands upon the capabilities of simple aggregates and provides user-defined precision guarantees through the use of mathematical models or sampling. In addition, representing time series as mathematical models, such as a polynomial regression model, provides functionality for data analysis such as interpolation or forecasting depending on the type of model utilized. AQP utilizing sampling reads only a subset of a time series and uses this subset to produce an approximate query result. A survey of mathematical models for representing time series can be found in the second chapter of [8], while additional information about sampling and integration of AQP into a DBMS can be found in [75]. For this survey we differentiate between approximation implemented as simple *aggregates* without guaranteed error bounds, and *AQP* utilizing either *mathematical models* or *sampling* to provide query results with precision guarantees.

**Stream Processing:** Describes the primary method each system supports, if any, for processing time series in the form of a stream from which data points are received one at a time. As each data point is processed independently of the full time series, each data point can be processed in-memory upon ingestion for applications with low latency requirements [76]. For queries with an unbound memory requirement or high processing time, AQP methods such as sampling or mathematical models can be used to lower the resource requirement as necessary [76]. Stream processing can be implemented using *user-defined functions*, as *functionality part of the TSMS*, or as a *query interface based on streams*. Examples of stream processing include online computation of aggregates, online constructing of mathematical models, removal of outliers in real-time, and realignment of data points in a limited buffer using windows. Operations performed after the data has been ingested by a TSMS, such as maintaining indexes or mathematical models constructed from data on disk, are not included. For a discussion of stream processing in contrast to traditional DBMSs see [76], while an in-depth introduction can be found in [10].

**Storage Engine:** The data storage component or system embedded into the application, used as external storage, or extended to enable storage of time series. If an existing system is used for data storage, we write the *name of the system*. For TSMSs were the storage component has been written from scratch, the column is marked as *implemented* with the implementation language added in parentheses.

**Storage Layout** The internal representation used by the system for storing time series, included due to the impact the internal representation has on the systems batch processing, stream processing and AQP capabilities.

TABLE 2
Summary of the Surveyed Systems in Terms of Functionality

| | Select | Insert | Append | Update | Delete | Aggregates | Join | Over | Analytics |
|---|---|---|---|---|---|---|---|---|---|
| *Internal Stores* | | | | | | | | | |
| tsdb [30] | ● | | ● | | ● | | | | |
| FAQ [31] | ● | | | | | ● | | | Jaccard, Top-K, AQP, etc |
| WearDrive [32], [33] | ● | | ● | | ● | | | ● | |
| RINSE [34], [35], [36] | ● | ● | ● | ● | | ● | | | NN-Search, AQP |
| Unnamed [37] | | | | | | ● | | | AQP |
| Plato [38] | ● | ● | ● | ● | ● | ● | ● | ● | Interpolation, AQP |
| Chronos [39], [40] | ● | ● | ● | ● | | ● | | | |
| Pytsms [41], [42] | ● | | ● | | | ● | | | |
| PhilDB [43] | ● | ● | ● | | | ● | | | Pandas |
| *External Stores* | | | | | | | | | |
| TSDS [44] | ● | | | | | ● | | | AQP |
| SciDB [45], [46], [47] | ● | ● | ● | ● | ● | ● | | | Linear Algebra, AQP |
| Respawn [48], [49] | ● | | ● | | | ● | | | |
| SensorGrid [50] | ● | | | | | ● | | ● | |
| Unnamed [51], [52], [53] | ● | | | | | ● | | | AQP |
| Tristan [54], [55] | ● | | ● | | | ● | | | AQP |
| Druid [56] | ● | | ● | | | ● | | | AQP |
| Unnamed [57] | ● | ● | ● | | | ● | | | AQP |
| Unnamed [58] | ● | | | | | ● | | ● | |
| Bolt [59] | ● | | ● | | ● | | | | AQP |
| Storacle [60], [61] | ● | | ● | | | | | | |
| Gorilla [62] | ● | | ● | | | | | | |
| Unnamed [63] | ● | | | | | ● | | | Frequency est, AQP, etc |
| servIoTicy [64], [65] | ● | ● | ● | ● | ● | ● | | | |
| BTrDB [66], [67] | ● | | ● | | | ● | | | |
| *RDBMS Extensions* | | | | | | | | | |
| TimeTravel [68], [69] | ● | ● | ● | ● | ● | ● | | | Forecasting, AQP |
| F²DB [70], [71] | ● | ● | ● | ● | ● | ● | | | Forecasting, AQP |
| Unamed [72], [73], [74] | ● | ● | ● | ● | ● | ● | | | Interpolation, AQP |

In addition to the classification criteria shown in Table 1, the TSMSs are summarized with regard to their supported query functionality in Table 2. To describe the functionality of the surveyed TSMSs with a uniform set of well-known terms, we elected to primarily express it using SQL. Two columns for keywords not part of SQL have been added to mark systems for which new data points can only be appended to an existing time series and to show which capabilities for data analytics each system supports. The full list of functionality included is: the capability to *select* data points based on timestamp or value, *insert* data points into a time series at any location, *append* data points to the end of a time series, *update* data points already part of a time series, *delete* data points from a time series, compute *aggregates* from a time series through the use of aggregate functions, *join* multiple time series over timestamp or value, perform computations *over* a time series using either window functions or user-defined functions, and any support for data *analytics* part of the system. In Table 2 we show a black circle if the TSMS is known to support that type of query through its primary API, and leave the cell empty if the TSMS is known to not support that type of query or if we were unable to determine if the functionality was supported or not. For the column *analytics* we list the supported functionality to differentiate the methods for data analytics supported by each system.

In the following three sections, we detail the primary contributions provided by each system in relation to the motivational use case and the classification criteria. Each section documents one set of systems based on the architecture of the system as shown in Table 1 and Table 2. In addition to a description of the contributions of each system, illustrations, redrawn from the original publications, are utilized to document the TSMSs. Two types of figures are used depending on the contribution of each system: *architecture diagrams* and *method illustrations*. Architecture diagrams use the following constructs: system components are shown as *boxes with full lines*, *lines* between components show undefined connections, data is shown as *boxes with rounded corners*, *arrows* show data flow between components, logically related components are grouped by *boxes with dotted lines* or by *dotted lines* for architectural layers, data storage is shown as *cylinders*, cluster nodes are components surrounded by *boxes with dashed lines*, and *text labels* are used as descriptive annotations. Method illustrations follow a less rigid structure due to the lack of correlation between the illustrated methods but in general layout of data in memory is shown as *boxes with full lines*. *Squares with rounded corners* are used for nodes in tree data structures. Constructs only used for a single figure are documented as part of the figure.

## 3 Internal Data Stores

### 3.1 Overview

Implementing a new TSMS as a single executable allows for deep integration between the data storage and data processing components. As the storage component is not accessible to other applications, the data layout utilized can be extensively optimized for the data processing component. Communication between the two components is also simplified as no communication protocol suitable for transferring time series is needed and no interface constrains access to the data storage layer unless an existing embeddable DBMS is used. The absence of external dependencies reduces the complexity of deploying the TSMS due to the system being self-contained. As a downside a system with an internal data store cannot utilize existing infrastructure such as a distributed DBMS or a DFS that already are deployed. In addition, if a new data storage layer is developed for a TSMS, instead of an existing embeddable DBMS being reused, time will need to be spent learning how that particular data store must be configured for it to perform optimally unless the TSMS provides automatic configuration and maintenance.

### 3.2 Systems

*tsdb* presented by **Deri et al.** [30] is a centralized TSMS designed for monitoring the large quantity of requests to the .it DNS registry. For storage tsdb utilizes the embeddable key-value store BerkeleyDB. The use of BerkeleyDB provides a uniform mechanism for storage of both the time series and the metadata used by the database. Data points ingested by tsdb are added to an array and when the array reaches a predefined size it is chunked and stored in BerkeleyDB. Each chunk is stored with a separate key computed by combining the starting timestamp for the array with a chunk id. tsdb requires all time series in the database to cover the same time interval, meaning each time series must have the same start timestamp and contain the same number of data points. This restriction simplifies computations that include multiple time series stored in the system, at the cost of tsdb not being applicable for domains where the need for regular time series adds complexity or is impossible to achieve. In terms of additional capabilities tsdb does not support stream processing, aggregation, or AQP.

The TSMS *FAQ* proposed by **Khurana et al.** [31] utilizes sketches organized in a range tree for efficient execution of approximated queries on time series with histograms as values. The system consists of multiple components as shown in Fig. 1. First, an index containing multiple types of sketches and histograms, each supporting a different type of query or providing a different trade-off between accuracy and query response time. An index manager and an error estimator are utilized by the query planner to select the most appropriate representation based on the provided query and the required error bound. However, utilization of the presented data structure, and thereby the TSMS in general, for use with stream processing of time series is relegated to future work. In addition, the external interface provided by the proof-of-concept implementation is not detailed in the paper.

*WearDrive* by **Huang et al.** [32] is a distributed in-memory TSMS for IoT, that demonstrates an increase in performance and battery life for a wearable device by transferring sensor data to a smartphone over a wireless connection. The system primarily uses volatile memory for storage as the flash storage used for wearable devices was identified as being a bottleneck in earlier work by Li et al. [33]. By extending the firmware used by a wearable device, WearDrive provides the same persistence guarantees as non-volatile flash storage. The system is split into two applications, as shown in Fig. 2, each built on top of a key-value store implemented for in-memory use. The store is organized as a sequential log of key-value pairs per application with each log file indexed by a hash map. WearCache is running on the wearable device and stores only a few sensor readings locally as the data is periodically pushed to remote memory or remote flash which is physically located on the smartphone, causing a drop in battery life for the smartphone. WearKV runs on the smartphone and stores the complete set of sensor readings collected from the wearable devices sensors in addition to the data sent from the smartphone itself to the wearable device. A simple form of stream processing is supported as applications can register to be notified when new values are produced by a specific sensor or for a callback to be executed with all the new sensor values produced over an interval. No support for AQP is provided by WearDrive. While the system is designed for a smaller scale, it follows a structure similar to other TSMSs with the wearable as resource-constrained device collecting data points. The data points are then transferred to the smartphone which serves as the system's backend and provides more capable hardware for storage and aggregate analysis.

*RINSE*, proposed by **Zoumpatianos et al.** [34], is a centralized system for data analytics supporting execution of efficient nearest neighbor queries on time series by utilizing the ADS+ adaptive index by Zoumpatianos et al. [35], [36]. The implementation is split into two components: a backend and a web frontend. The backend serve as the data storage and indexing component, storing time series in an unspecified ASCII on-disk format indexed using ADS+. The web frontend is served through NodeJS and provide the means to execute nearest neighbor queries by drawing a pattern to search for. In addition, the capabilities of the data storage component are available through a TCP socket. The use of ADS+ for its index provides the system with multiple capabilities. First, as ADS+ is an adaptive tree-based index only internal nodes are initialized while the leafs, containing the data points from the time series, are only initialized if that part of the time series is used in a query. This reduces the time needed to construct the index before queries can be executed compared to alternative indexing methods. Second, approximate queries can be executed directly on the index, providing both AQP and an index for exact queries using the same data structure.

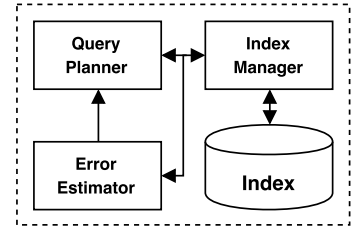

Fig. 1. The architecture of the FAQ system, redrawn from [31]

A centralized TSMS that utilizes models for AQP was proposed by **Perera et al.** [37]. The system manages a set of models for each time series and the TSMS optimizer selects the most appropriate model for each query, falling back to the raw data points if necessary. To construct the models the system splits each time series into smaller segments and attempts to approximate each segment with a model. The use of segmentation allows different models optimized for the structure of each segment to be used. If a model with the necessary precision cannot be constructed for a segment then the segment is represented only by the data points, meaning only exact queries can be executed for that segment. The authors proposed an extension of the SQL query language with additional capabilities for AQP: support for specifying which model to use, the maximum error bound for the query, and a maximum query execution time. How a particular query should be executed is left to the query optimizer which given the provided model, maximum query execution time, the required error bound and statistics collected from past queries determines if a model can be used or if the query must be performed using the raw data. However, in the current R based implementation, AQP using models is only supported for SQL aggregate queries. A later paper by the same authors, Perera et al. [77], reused the method of approximating time series using models to reduce the storage requirements of materialized OLAP cubes.

Another centralized TSMS that implements model-based AQP is *Plato* proposed by **Katsis et al.** [38]. The system combines an RDBMS with methods from signal processing to provide a TSMS with functionality for data analytics, removing the need for exporting the data to other tools such as R or SPSS. Plato consists of three layers, as shown in Fig. 3, in order to accommodate implementation of models, construction of model-based representations of time series, and querying of time series represented as models. At the lowest level, an external interface allows for additional models to be implemented by domain experts. By providing an interface for implementing new models, Plato becomes simple to extend for new domains. The implemented models, however, have to be constructed manually by a database administrator by fitting a model to a specific data set, as an automated model selection through a cost function is left as future work. AQP is supported by querying the models manually through one of two extensions of SQL. ModelQL is designed to appeal to data analysts familiar with R or SPSS, while the InfinityQL language is designed to be familiar to users experienced with SQL. Queries are evaluated directly on a model if the model implements the functionality being used in the query. If the functionality is not implemented, the model is instantiated at the resolution necessary for the particular query. As future work the authors propose that models are not used only as a method for representation of time series, but also returned as the query result in order to provide insight into the structure of the data.

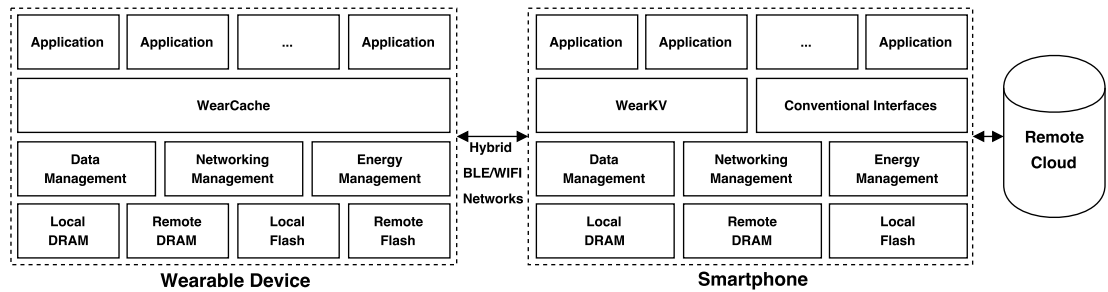

Fig. 2. The split architecture of WearDrive, remote resources are logical and actually located on the other device. The figure was redrawn from [32]

The centralized open-source system *Chronos* by **Chardin et al.** [39] is designed as a TSMS for monitoring industrial systems located in a hydroelectric dam. Due to the location, all persistent storage is flash based to provide increased longevity, with a substantial drop in write performance compared to spinning disks. To accommodate the use of flash memory, an abstraction layer, proposed by Chardin et al. [40], is implemented on top of the file system so writes are kept close to the previous write on the flash memory ensuring near sequential writes. This is preferable as write duration increases the further from the previous write it is performed. The abstraction layer index data stored on this logical file system through a B-Tree implementing a new splitting algorithm modified to not degrade in performance when used with flash memory. The system provides a simple form of stream processing through efficient support for out-of-order inserts by buffering and reorganizing ingested data before making it persistent, thereby providing a trade off between efficient out-of-order inserts and the possibility of data loss due to hardware failure as the temporary buffers are stored in volatile memory. However, no support for stream processing using user-defined functions, functionality for construction of aggregates or creation of approximate representation of time series are provided.

*Pytsms* and its extension *RoundRobinson* serve as the reference implementation for two formalisms for TSMSs proposed by **Serra et al.** [41], [42]. However, as the implementation only serves as the proof-of-concept for the formalisms presented in the papers, no attempt was made to make them efficient nor make the implementation comparable in functionality to existing TSMSs. Pytsms implements a centralized TSMS for storage of time series. In addition to storage, Pytsms implements a set of operations that can be performed on time series. RoundRobinson provides a multiresolution abstraction on top of Pytsms. The concept of multiresolution time series is implemented as buffers and round robin discs. When a buffer is filled, either through a time series being bulk loaded or streamed from a sensor into Pytsms, an aggregate is computed by executing an aggregation function with the data points in the buffer as input. The aggregate is added to the disc which functions as a fixed size round robin buffer discarding old aggregates as new are added. As both buffers and discs can be configured, the system is capable of creating any aggregated view of a time series that is needed such as high resolution aggregates for the most recent part of a time series and then decrease the resolution in increments for historical data. An example of such a configuration is shown in Fig. 4. The example shows a schema for a time series represented at different resolutions where only a few aggregates are stored for the entire six hundred days the time series represents, while a high number of data points are stored for the discs containing more recent data.

*PhilDB* proposed by **MacDonald** [43] is an open-source centralized TSMS designed for data analytics in domains where updates to existing data points are a requirement. To preserve the existing readings and allow for old versions of a time series to be analyzed, PhilDB separates storage of time series into two parts: data points and updates. The data points are stored sequentially in binary files as triples containing a timestamp, a value, and last an integer used as a flag to indicate properties of the data point, for example if the data point is missing a value in an otherwise regular time series. Any updates performed to a time series is stored in an HDF5 file collocated with the binary file storing the data points. Each row in the HDF5 stores the same data as the triples in the binary file with an additional timestamp added to indicate when the update was performed. As the original data points are kept unchanged, queries can request a specific version of the time series based on a timestamp. Additionally, a time series is identified by a user-defined string and can be associated with additional attributes such as the data source for the time series, all stored using SQLite.

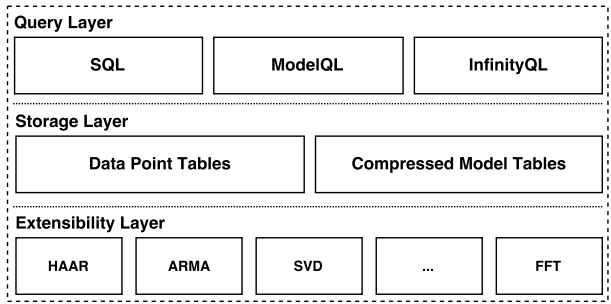

Fig. 3. The architectural layers of the Plato TSMS, redrawn from [38]

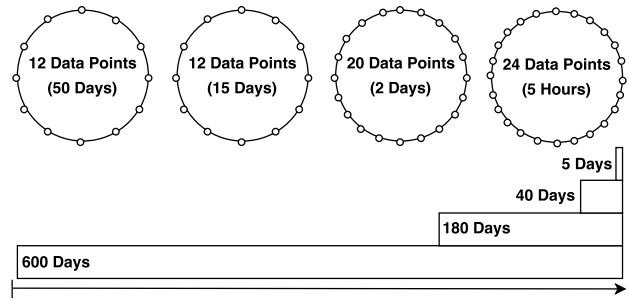

Fig. 4. A time series stored at multiple resolutions using discs of static size and with static granularity for each measurement, redrawn from [41]

Reading and writing the time series to disk is implemented as part of PhilDB, however, as its in-memory representation it uses Pandas. The use of Pandas provides a uniform data format for use by the DBMS and allows it to interface directly with the Python data science ecosystem.

### 3.3 Discussion

A few common themes can be seen among the TSMSs that use an internal data store. First, only WearDrive [32] is distributed and is specialized for use with wearable devices. Instead, researchers implementing distributed TSMSs use existing distributed DBMSs and DFSs running externally from the TSMS as will be described in Section 4. Similarly, only tsdb [30], Plato [38] and PhilDB [43] are intended for use as general TSMSs and have implementations complete enough for this task based on the description in their respective publications. The remaining systems serve only as realizations of new theoretical methods, FAQ [31] is used for evaluating a method for model-based AQP, RINSE [34] demonstrates the capabilities of the ADS+ index [35], [36], Chronos [39] illustrates the benefits of write patterns optimized for flash storage but does neither provide thread-safety nor protection against data corruption, the system by Perera et al. [37] is incomplete, and both Pytms and RoundRobinson [41], [42] are simple implementations of a formalism. In summary, researchers focusing on development of general purpose TSMSs are at the moment focused on systems that solve the problem at scale through distributed computing, with centralized systems being relegated to being test platforms for new theoretical methods. A similar situation can be seen in terms of AQP and stream processing as none of the three general purpose TSMS, tsdb [30], Plato [38] and PhilDB [43], support stream processing, and only Plato implements model-based AQP. In addition, Plato provides an interface for users to implement additional models, making it possible for domain experts to use models optimized for a specific domain. For the research systems, FAQ [31], RINSE [34] and the system Perera et al. [37] all support AQP using models, while Pytms and RoundRobinson [41], [42] only provide user-defined aggregates. None of the four research systems implemented functionality for stream processing.

## 4 EXTERNAL DATA STORES

### 4.1 Overview

Implementing a new TSMS by developing a data processing component on top of an existing DBMS or DFS provides multiple benefits. Existing DBMS and DFS deployments can be reused and knowledge about how to configure the system for a particular workload can still be applied. The development time for the system is also reduced as only part of the TSMS must be implemented. In terms of disadvantages, the choice of DBMS will restrict how data is stored and force the processing component to respect that data model. Second, deployment of the system becomes more complicated as multiple separate systems must be deployed, understood, and configured for a particular workload.

### 4.2 Systems

*TSDS* developed by **Weigel et al.** [44] is an open-source centralized system designed for data analytics. TSDS supports ingesting multiple time series stored in different formats, caching them and then serving them from a central location by providing a HTTP interface for querying the time series. As part of TSDS a TSMS named TSDB is developed. TSDB is used primarily for caching and storing each separate time series in a sequential binary format on disk to reduce query response time. Multiple adapters have been implemented to allow for ingesting times series from sources such as ASCII files and RDBMSs. Transformations through filters and sampling can be performed by TSDS so only the time interval and resolution requested are retrieved from the system without depending on an external data analytics program such as R or SPSS. As the system is designed for ingesting static data sets no methods for stream processing is provided. Development of TSDS is no longer in progress, however, the TSDS2 project is developing a new version of the TSMS.

Proposed by the **SciDB Development Team** [47] and **Stonebraker et al.** [45], [46], *SciDB* is a distributed DBMS designed to simplify large scale data analytics for researchers. While not designed exclusively for use as a TSMS, SciDB includes functionality for storing sequential data from sensors used to monitor experiments. SciDB stores data as N-dimensional arrays instead of tables like in an RDBMS, thereby defining an implicit ordering of the data stored. The arrays are versioned and updates to existing arrays produce a new version. A SciDB cluster uses an instance of PostgreSQL for storing the system catalog containing configuration parameters and array metadata, including for example the array version number. Two query languages are implemented: AQL a language similar to SQL that provides a high level declarative interface and AFL which is a lower level language inspired by APL, which allows a chain of operators to be defined directly. Matrix operations are performed outside SciDB using the ScaLAPACK linear algebra library executing alongside the DBMS. In addition, multiple client libraries have been developed, such as SciDB-R that supports R programs executed in SciDB. Support for AQP is provided through a function for sampling a subset of data from arrays, and stream processing is limited to an API inspired by Hadoop Streaming for streaming arrays through external processes akin to Unix pipes. Other DBMSs have been implemented using the open-source SciDB code base. As an example Li et al. [78] created FASTDB, a DBMS specialized for use in astronomy. FASTDB extends SciDB in several areas: an interactive frontend for data exploration, additional monitoring for cluster management, a parallel data loader, a method for dynamically determining the appropriate size of chunks when splitting the data into chunks for storage, and last enhancements of the query optimizer in SciDB was implemented. For additional information about DBMSs based on arrays see the survey by Rusu et al. [79].

The distributed TSMS *Respawn* presented by **Buevich et al.** [48] is designed for monitoring using large scale distributed sensor networks and provide low latency range queries on the produced data points. In Respawn, data storage is performed by the multi-resolution TSMS Bodytrack DataStore which is integrated into two different types of nodes: sensor network edge nodes and cloud nodes as shown in Fig. 5. Edge nodes are ARM based devices that are placed at the edge of a sensor network and ingest the data points produced by the sensor network and compute aggregates at multiple different resolutions, enabling reduced query response time for queries at resolutions lower than what the sensor is being sampled at. In addition to the distributed edge nodes, server grade cloud nodes are deployed to provide a cache of the data stored in the edge nodes, with segments of time series being migrated using two preemptive strategies. Periodic migration forces migration of low resolution aggregates at specific intervals, as the high resolution data usually is used only if an unusual reading is found through analysis of the low resolution time series. Proactive migration defines how segments of a time series should be migrated based on the standard deviation of the data points, as this can indicate irregular readings worth analyzing further. Queries are performed through HTTP requests and to determine if they should be sent to an edge node or to a cloud node a dispatcher is implemented to track what data has been migrated. After the location of the time series at the requested resolution has been provided, all communication is done directly between the client and the edge node or cloud node, without routing it through the dispatcher. Respawn was utilized as part of the Mortar.io platform for building automation presented by Palmer et al. [49]. Mortar.io provides a uniform abstraction on top of the automation infrastructure, such as sensors, providing a web and mobile interface in addition to simplifying the development of applications communicating with the distributed hardware components.

*SensorGrid* is a grid framework for storage and analysis of sensor data through data analytics proposed by **Cuzzocrea et al.** [50]. The proposed architecture consists of sensors, a set of stream sources that represent nodes for interacting with sensors and transition of the collected data points to stream servers that store the data. To reduce query response time for aggregate queries, the stream servers pre-compute a set of aggregates defined by the system with no mechanism provided for defining a user-defined aggregation function. The aggregates are computed using an application-specific time interval and the sensor that produced the data points, with the requirement that an aggregation hierarchy for sensors must be explicitly defined. Distributed processing is implemented as part of the stream servers providing them with three options when a query requests data not present at a node: redirect the query, use a local approximation of the data requested by the query at the cost of accuracy, or decompose the query into multiple sub queries and send them to other nodes in the grid. In addition to approximation, the system supports execution of queries over fixed windows, and even continuous queries over moving windows using SQL for stream processing. The SensorGrid architecture was realized as a distributed TSMS and used for hydrogeology risk analysis developed at IRPI-CNR, where the stream source and stream server nodes are integrated with the RDBMS Microsoft SQL Server 2000 and a web interface for visualizing the data.

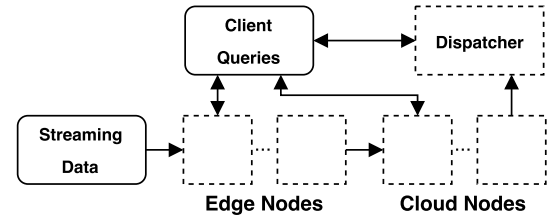

Fig. 5. Architecture of the Respawn TSMS with edge nodes collecting data at the source and migrating it to cloud nodes, redrawn from [48]

**Guo et al.** [51], [52], [53] proposed a TSMS that uses mathematical models stored in a distributed key-value store to reduce query response time. The system consists of three parts, a key-value store for storing segments of the time series represented by models, two in-memory binary trees used as an index for the modelled segments, and last an AQP method based on MapReduce. To enable the index to support both point and range queries, the binary trees are ordered by intervals, with one tree constructed for time intervals and another for value intervals. Similarly, two tables are created in the key-value store as one stores the modelled segments with the start time and end time as part of each segment's key, while the other table stores each model with the minimum and maximum value that is part of the modelled segment as part of its key. Query processing is shown in Fig. 6 and is performed by first traversing the index to lookup segments relevant to the query by determining which nodes overlap with the time and values specified in the query. Second, MapReduce is used to extract the relevant segments from the key-value store and re-grid them. As each node in the tree might reference multiple segments, the mapping phase prunes the sequence of proposed segments to remove those that do not fit the provided query, and last the reducer phase re-grids the segments to provide values approximating the original data points. The described method does only rely on the current data point and a user-specified error bound

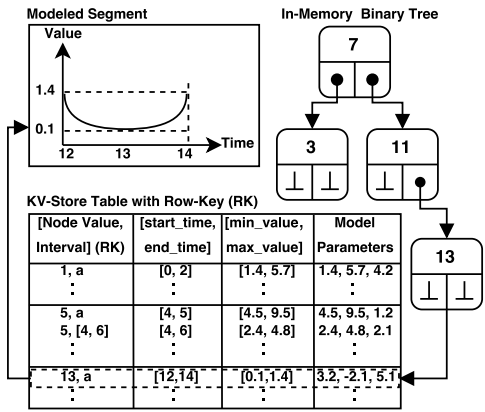

Fig. 6. Model-based AQP combining an in-memory tree and a distributed KV-store for the TSMS by Guo et al. [51], [52], [53], redrawn from [51]

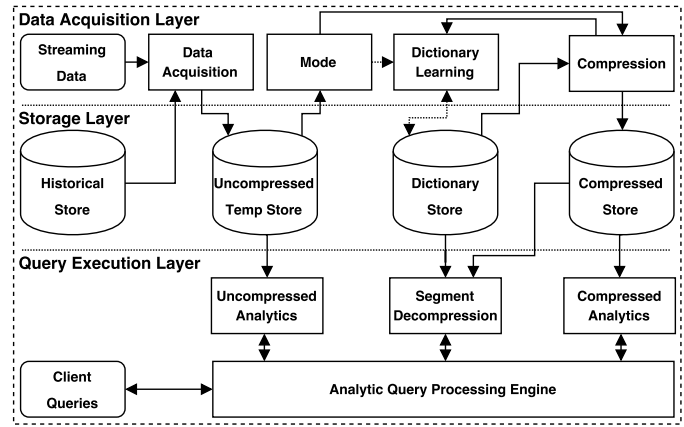

Fig. 7. The MiSTRAL architecture realized by the TSMS Tristan with data flow occurring only in offline mode as dotted arrows, redrawn from [55]

allowing it to be implemented in terms of stream processing.

*Tristan* is a TSMS designed by **Marascu et al.** [54] for efficient analytics of time series through the use of AQP and online dictionary compression. The system is based on the MiSTRAL architecture also proposed by Marascu et al. [55]. The system's architecture is shown in Fig. 7 and is split into three different layers: a data acquisition and compression layer, a storage layer, and last a query execution layer. The data acquisition and compression layer ingests data into segments which are then stored in a temporary data store managed by the storage layer. When a segment reaches a pre-configured size, the segment is compressed as a sequence of smaller fixed size time series using a dictionary provided as a parameter to the system. The accuracy of the compressed time series can be configured based on the number of fixed sized time series it is represented as. Creation of the dictionary is performed through offline training, however, when operational, the system will continue to adjust the dictionary based on predefined time intervals. The storage layer manages access to the uncompressed time series, the temporary store, the dictionaries used for compression and the compressed segments. The current implementation of the storage layer is based on a customized version of the open-source in-memory DBMS HYRISE. Last, the query execution layer receives queries from the user in an unspecified manner and format. Tristan is capable of using three different methods for answering the query: if the query is for data that is in the process of being compressed, the query is executed against the temporary data store, if the query references a compressed segment Tristan determines if the query can be answered directly from the compressed representation and only decompresses the time series segment if answering the query from the compressed form is not possible. The system supports use of AQP to reduce query response time due to its functionality for selecting the precision for compressing segments by adjusting how many smaller fixed size time series it is represented by. Tristan only supports stream processing through its capability to transform incoming data into an approximate model by buffering it in an uncompressed form in a temporary data store. No support for transforming the data through user-defined functions are possible.

**Yang et al.** implemented the open-source system *Druid* [56] as a distributed TSMS for efficiently ingesting time series in the form of events from log files, and then performing OLAP-based data analytics on the ingested data. Druid is based on a shared nothing architecture coordinated through Zookeeper in-order to optimize the cluster for availability. Each task required by a Druid cluster is implemented as a specialized type of node as shown in Fig. 8. Real-Time nodes ingest events and store them in a local buffer to efficiently process queries on recent events in real-time. Periodically a background task collects all events at the node and converts them into immutable segments and uploads them to a DFS. Historical nodes execute queries for segments stored in the DFS and caches the segments used in the query to reduce query response time for subsequent queries. Coordinator nodes manage the cluster's configuration stored in a MySQL database, in addition to controlling which segments each historical node must serve. Last, the broker nodes accept queries formatted as JSON through HTTP POST requests and route them to the relevant real-time and historical nodes. When the query result is returned it is cached at the broker for subsequent queries. Druid does not provide any support for stream processing, and the authors suggest combining Druid with a stream processing engine. However, Druid does support aggregation of time series when ingested or when queried, and provides a set of aggregate functions while also supporting user-defined aggregation functions in addition to HyperLogLog for approximate aggregates.

**Huang et al.** [57] designed a distributed TSMS that uses IBM Informix for storage and provides a uniform SQL interface for querying both time series and relational data together. The system is classified as an Operational Data Historian (ODH) by the authors, marking monitoring as a clear priority. The TSMS consists of three components: A configuration component, a storage component, and a query component as shown in Fig. 9. The configuration component manages metadata for use by other components in the system. Concretely, it provides information about the data sources the system is ingesting data from, and the IBM Informix instances available to the system. The storage component ingests data through a set of writer APIs into an appropriate data structure and compresses the data before it is stored. Different data structures are used depending on if the time series is regular or irregular. The three supported data formats are shown in Fig. 10. The first two formats store a single time series and contain four components, a starting timestamp, a device id, a counter and a blob of values. If the time series consists of data points separated by regular intervals using the value from the data points are stored without timestamps using the first format. If the time series is not regular then data points are stored as both the delta time compared to the starting timestamp and values using the second format. Last using the third format, a group of time series can be combined and stored in a single data structure, in which case the device id is changed to a group id and each element in the blob will then contain a reduced device id, a time delta and a value. To compress time series segments two existing compression algorithms are used, both supporting either lossless compression or lossy compression within an error bound. For stable values the data is compressed using a linear function, while quantization is used for fluctuating time series. The query component provides a virtual table as the query interface for the time series, allowing SQL to be used as a uniform query interface for both time series and relational data stored in the same TSMS.

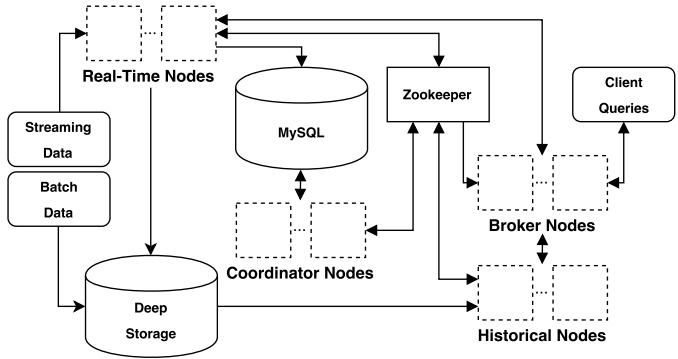

Fig. 8. The architecture of a Druid cluster annotated with information about data flow between nodes in the cluster, redrawn from [56]

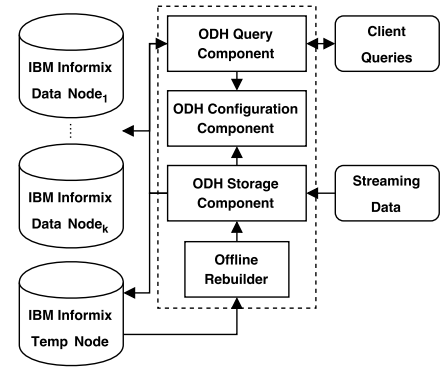

Fig. 9. Architecture of the TSMS proposed as an Operational Data Historian (ODH) by Huang et al. [57]. The figure was redrawn from [57]

**Williams et al.** [58] propose a distributed system for ingesting and analyzing time series produced by sensors monitoring industrial installations. The system is based on Pivotal's Gemfire in-memory data grid due to the authors arguing that a disk-based system is an unsolvable bottleneck for data analytics. The data points produced by the sensors are ingested by a proprietary processing platform that cleans the data and performs real-time analytics before the data is inserted into the in-memory data grid. In the in-memory data grid, the cleaned data points are organized into a set of bins each containing a fixed time interval, sorted by time and identified by both the sensor id and the id of the machinery the sensor is installed on. The authors argue that each bin should only contain a few minutes of data as a compromise between duplicating the machine and sensor identifier, while also reducing the amount of data being retrieved during query execution to extract a subset of the data points in a bin. Inside a bin, the segment of data points produced by a specific sensor is stored as a doubly linked list, allowing memory usage to scale with the number of data points in the bin while achieving comparable read and write performance to a statically allocated circular array. Due to memory limitations, the in-memory data grid is used as a FIFO cache, storing all data points over a pre-specified time frame, with efficient transfer of the data to a long term disk based solution designated as future work. Additionally, neither AQP nor pre-commutation of aggregates when the data is ingested is supported by the system. For more information about in-memory big data storage and data analytics see the survey by Zhang et al. [80].

*Bolt* developed by **Gupta et al.** [59] is a distributed open-source TSMS designed for being embedded into applications to simplify the development of applications for IoT. Data is organized into tuples containing a timestamp and a collection of tag and value pairs, providing the opportunity to store not only readings from sensors but also annotate each reading with metadata. Bolt's implementation is based on encrypted streams in a configuration with one writer, the application that created the stream, and multiple readers that can request readings from the stream. Sample-based AQP is provided as a function call with a parameter indicating how many elements should be skipped between each returned sample. The streams are chunked and stored sequentially in a log file on disk, while an index is kept in memory to provide efficient queries for tags or tags in a specific time frame. If the index reaches a pre-specified size, it is spilled to disk with only a summary kept in-memory, making querying historical data more expensive than recent data points. Sharing data between multiple instances of Bolt is achieved by having a metadata server provide information about the location of available streams, while exchange of the encrypted data is performed through either Amazon S3 or Microsoft Azure.

A similar TSMS *Storacle*, was developed for monitoring a smart grid and proposed by **Cejka et al.** [60]. The system is designed to have low system requirements allowing it to be used throughout a smart grid and connected to local sensors and smart meters. The TSMS provides the means for computing aggregates and in general process data from multiple sources locally before it is transmitted to the cloud over SSH. The cloud is defined as a remote storage with support for replication, offline data processing and remote querying. Storacle uses protocol buffers as its storage format and uses a three-tiered storage model consisting of in-memory storage, local storage and cloud storage. Storacle supports multiple parameters for configuring the amount of data that should be kept in-memory or stored locally for efficient access. However, the most recent data is kept when data is transferred to the next tier, ensuring that the most recent readings always are available. In addition to immutable storage of time series Storacle provides support for mutable storage of tags and meta-fields, both of which are lists of strings but tags can be used as part of a query while meta-fields cannot. Last, Storacle uses stream processing to process each ingested data point and produce aggregates from the data such as the number of data points, the average of their values, the min and max values, in addition to a histogram of observed values. Additional software built on top of Storacle is presented by the authors in the form of a CSV exporting tool and a tool that continuously retrieves the latest data points and computes aggregates for monitoring. In addition to the application presented in the original paper, Faschang et al. [61] proposed an integrated framework for active management of a smart grid in which Storacle was integrated with a message bus named GridLink.

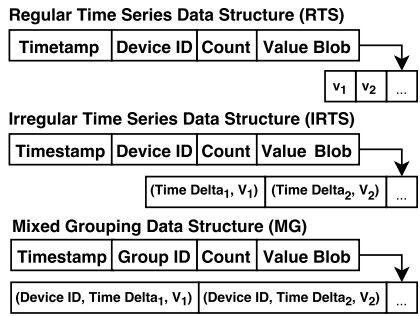

Fig. 10. The data structures for storage of time series used in the TSMS proposed by Huang et al. [57]. The figure was redrawn from [57]

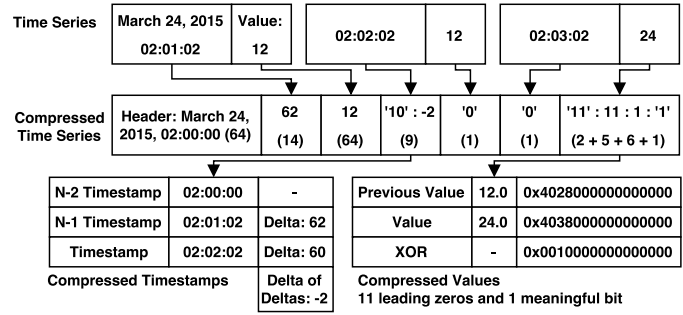

Fig. 11. Gorilla's compression with bit patterns written in single quotes and the size of values in bits written in parentheses, redrawn from [62]

**Pelkonen et al.** designed *Gorilla* [62] at Facebook as a distributed in-memory TSMS to serve as a caching layer for an existing system based on Apache HBase for monitoring the infrastructure at Facebook. Gorilla was designed to reduce query response time for data points collected in the last 26 hours, with the time frame determined by analyzing how the existing system was used. The system was designed as a cache instead of a replacement as the existing TSMS based on HBase contained petabytes of data. Data points ingested by Gorilla contain three components: a key, a timestamp and a value. The key is unique for each time series and used for distributing the data points, ensuring each time series can be mapped to a single host. Each time series is stored in statically sized blocks aligned with a two-hour window, with one block being written to at a time per time series and older blocks being immutable. Gorilla uses a lossless compression method for both timestamps and values based on the assumption that data points are received at an almost regular interval as shown in Fig. 11. The method works by having each block be prefixed with a timestamp at full resolution, and the first data point stored with the timestamp encoded as the delta between the block's prefixed timestamp and the data point's timestamp, and the data point's value stored in full. Subsequent data points are compressed using two methods, one for timestamps and another for values. Timestamps are compressed by first computing the deltas between the current and previous timestamps and then storing a delta of these deltas using variable length encoding. Values are compressed by first XOR'ing the current and previous value, and then storing the difference with zero bits trimmed if possible. Both methods resort to storing a single zero bit in the instance where the computed delta or XOR result show no difference. To ensure availability Gorilla replicates all data to a different data center but provides no consistency guarantees. As each query is redirected to the nearest location the TSMS trades lower query response time for consistency. For fault tolerance Gorilla achieves persistence by writing the time series data to a DFS with GlusterFS being used at the time the paper was written. Data processing and analytics are performed by separate applications that retrieve blocks of compressed data from the in-memory TSMS through a client library, with applications for computing correlation between time series, real-time plotting and computation of aggregates being developed. Similarly, no capabilities for stream processing are implemented directly as part of Gorilla. An open-source implementation of the ideas presented in the paper was published by Facebook as the Beringei project.

**Mickulicz et al.** [63] propose a TSMS for data analytics using a novel approach for executing approximate aggregate queries on time series with the storage requirement defined by the time interval and not the size of the raw data points. By defining what aggregate queries will be executed, what granularity aggregates will be required in terms of time, and with what error bound, a hierarchy of aggregates can be computed during data ingestion. The computed aggregates are stored in a binary search tree with the aggregates spanning the smallest time intervals at the leaves and an aggregate over the entire time series at the root. This structure enables efficient query response time for aggregate queries with differently sized time intervals, as aggregates over large time intervals can be answered using aggregates at the root of the tree, while queries over smaller time intervals use aggregates from the leaves. The presented approach is implemented as part of a distributed TSMS used for analyzing time series of events from mobile applications. The system consists of three components: a set of aggregation servers that computes multiple aggregates based on the ingested events, multiple MySQL RDBMSs for storing both the unmodified events and the aggregates computed, and last a querying interface routing queries to the relevant RDBMS. In terms of aggregates, the system uses a simple sum for counting events, HyperLogLog for approximately computing the number of distinct values, and the Count-Min Sketch for approximate frequency estimation.

*servIoTicy* is a TSMS implemented by **Pérez et al.** [64] for storing time series and metadata from IoT devices and is split into a frontend and a backend component. The frontend provides a REST API serving JSON for interacting with the system. To increase the number of devices that can communicate with the system, the REST API is accessible using multiple different communication protocols. The backend provides data storage using a combination of Couchbase and Elasticsearch for storage and query processing. Data is stored in Couchbase as one of two differently structured JSON documents. The first JSON format is used for storing metadata about the IoT devices the system is ingesting data from, while the second format is used to store data received from each IoT device. To reduce query processing time, the data stored in Couchbase is indexed by Elasticsearch. Stream processing using user-defined topologies is implemented through integration with Apache Storm. Apache Storm has also been extended with a mechanism for updating a function being executed in an Apache Storm Bolt. The system can then execute a specific version of the Bolt depending on the data points being processed. To change the topology it is still required that the system is terminated before a new topology can be deployed. In terms of AQP no support is currently provided by servIoTicy. Due to servIoTicy being part of the COMPOSE project [81], it was later integrated with the web service discovery system iServe, to augment the data stored in servIoTicy as documented by Villalba et al. [65].

*BTrDB* proposed by **Andersen et al.** [66], [67] is an open-source distributed TSMS designed for monitoring high precision power meters producing time series with timestamps at nanosecond resolution. A new system was developed as existing TSMSs were evaluated and determined to be unable to support the required ingestion rate and resolution of the timestamps. In addition to the high resolution no guarantees can be made about the ordering of data points or if each data point only will be received once. In BtrDB both problems are solved by storing each time series as a copy-on-write k-ary tree with each leaf node in the tree containing data points for a user-defined time interval and each internal node containing aggregates such as min, max, and mean of the data points stored in children nodes. An example of the tree is shown in Fig. 12. This data structure provides multiple benefits. Data points received out of order can be inserted at the correct leaf node without any changes to the rest of the stored time series, support for efficient queries at multiple different resolutions is also enabled by the tree structure, as a time series can be extracted at multiple different resolutions by the pre-computed aggregates, and last as the tree is a persistent data structure, older versions of a time series are available despite new data points being inserted and other being deleted. The system is split into three components. BTrDB provides the query interface, optimization of data ingestion through buffering, and manipulation of the k-ary trees. A DFS in the form of CEPH is used for persistent storage of the k-ary trees constructed for each time series, and last MongoDB is used to store the metadata necessary for BTrDB. To reduce the storage requirements for the trees, compression is applied. To compress a sequence, the compression algorithm computes the delta to the mean of the previous deltas in the sequence, and the computed deltas are then encoded with Huffman encoding. Apart from the API being structured around streams BTrDB does not provide any stream processing capabilities. However, the low query responds time for reads and writes provided by the TSMS were utilized to implement the DISTIL stream processing framework that integrates with BTrDB [67].

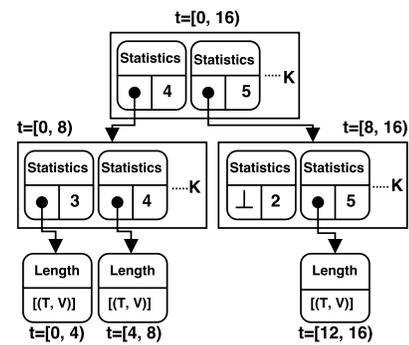

Fig. 12. The copy-on-write K-ary tree stores time series in the leafs while aggregates and version numbers are in non-leaf nodes, redrawn from [66]

### 4.3 Discussion

The main domains covered by systems with external data stores are IoT, monitoring of industrial or IT systems, and management of scientific data. The developed systems are in general complete, in use, and support larger data sets by scaling out through distributed computing and caching of data in memory for efficient processing of either the most recently collected or queried data. TSDS [44] and Tristan [54] are exceptions as both of them operate as centralized systems limiting their scalability. An interesting observation is also that none of the systems implement an entirely new data storage layer, and instead to varying degrees reuse existing systems for example MySQL, HDFS, and Couchbase. Only three systems, the system by Williams et al. [58], Gorilla [62], and servIoTicy [64], provide no capabilities for constructing approximate representation of time series while the remaining systems support either simple aggregates or AQP. The systems implementing AQP using models are either research systems or provide limited functionality for implementing additional user-defined models. The system by Guo et al. [51], [52], [53] is used to demonstrate a new indexing and query method and provides a generic schema for storing the models. Tristan employs AQP by representing time series using multiple smaller fixed sized time series but the TSMS provides no interface for implementing other methods for representing time series suitable for AQP. Druid [56] allows implementation of aggregation methods through a Javascript scripting API or through a Java extensions API. The TSMS developed by Huang et al. [57] provides no mechanisms for implementing new models for use by the system and only two types of models are implemented and used for compression. No interface is documented for extending the system by Mickulicz et al. [63] with new models, however, the authors note that the approximate representation used as part of the proposed tree structure can take different forms depending on the queries the tree the must support. In summary, only Druid [56] and to a certain degree the system by Mickulicz et al. [63] provide an interface for end users to implement alternative representations of time series so domain experts can use models optimized for a particular domain. Multiple aspects of stream processing are utilized by this category of systems. Stream processing using user-defined computation is provided by SensorGrid [50] through the use of SQL window functions, while servIoTicy [64] and the TSMS by Williams et al. [58] support user-defined functions. Tristan [54], Guo et al. [51], [52], [53], Huang et al. [57], and Mickulicz et al. [63] construct models from data points online. Last, SciDB [45], [46], [47], Bolt [59] and BTrDB [66], [67], logically structure some APIs as streams.

# 5 RDBMS EXTENSIONS

## 5.1 Overview

Existing RDBMSs have been extended with native functionality for efficient storage, querying and analysis of time series stored in the RDBMS. Implementing functionality for processing time series data directly into an RDBMS provides multiple benefits. It simplifies analysis by providing storage and analysis through one system, removing the need for data to be exported and the analysis performed using an additional program such as R or SPSS. Existing functionality from the RDBMS can be reused for the implementation of the extensions. Last, extending a known system, such as PostgreSQL, allows knowledge of said system to be reused. However, as with all extensions of existing systems, decisions made about the RDBMS implementation can restrict what changes can be made, and functionality such as transactions add unnecessary overhead if the time series is immutable.

## 5.2 Systems

*TimeTravel*, by **Khalefa et al.** [68], extends PostgreSQL with the capability to do model-based AQP and continuously compute forecasts for time series stored in the centralized RDBMS. Using forecasting and an extended SQL query language, TimeTravel simplifies data analytics by providing a uniform interface for both exact query processing used for historical values and approximate query processing used for historical and future values. Use of AQP is necessary as the computed forecasts will have some estimation error due to the exact values being unknown. However, the use of AQP is also beneficial for reducing query time on historical data. The architecture of TimeTravel can be seen in Fig. 13, and consists of a hierarchical model index, offline components for building and compressing the model index, and online components for query processing and maintenance of the model index. The models range from a coarse grained model with a high error bound at the top to multiple finer grained models with lower error bounds at the bottom. This hierarchy allows the system to process queries more efficiently by using a model of the right granularity based on the required accuracy of the result. To build the model hierarchy, the system requires multiple parameters to be specified: First, hints about the time series seasonal behavior, second, error bound guarantees required for the application that will query the data, and last what method to use for forecasting. The system builds the hierarchical model index by creating a single model, and then based on that model's error, it splits the time series into more fine-grained models based on the required error bound and provided seasonality. In terms of stream processing, TimeTravel provides no extensions to PostgreSQL other than the module for maintaining the model index automatically. The methods created as part of TimeTravel have been incorporated into a Electricity Data Management System (EDMS) developed as part of the MIRABEL smart grid project [82] by Fisher et al. [69]. To facilitate both exact and approximated analysis of time series in the EDMS, a real-time data warehouse is implemented as part of the system. For approximate queries the AQP functionality implemented as part of TimeTravel were used while exact queries were answered from the data points stored in an array in PostgreSQL.

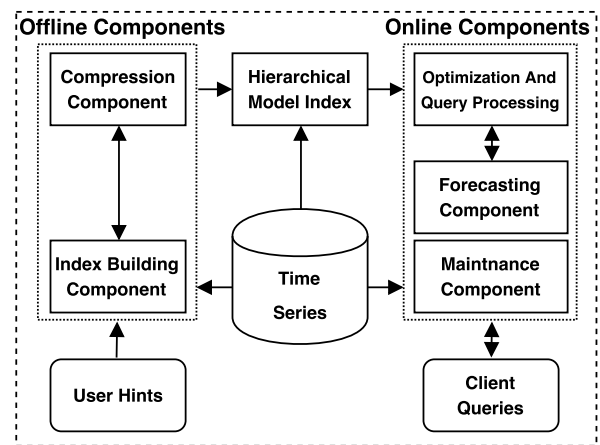

Fig. 13. Architecture of the TimeTravel TSMS with components for index construction and query processing. The figure was redrawn from [68]

Another endeavor to extend PostgreSQL with the capability to forecast time series in a centralized RDBMS is $F^2DB$ proposed by **Fischer et al.** [70]. The use case for $F^2DB$ is data analytics using data warehouse based business intelligence, and the system extends the SQL query language with statements for manually creating models and performing forecast on said models. As an alternative to creating a model by specifying it as part of an SQL statement, $F^2DB$ can compute which model provides the best fit for a given time series based on an evaluation criteria. The system provides a general architecture for implementing forecast models, allowing domain experts to add additional models to the system so they can be used through the SQL query language. The current implementation provides the means to create models based on ARIMA and exponential smoothing. All models are stored in a central index and maintained automatically by the system as shown in Fig. 14, but no extensions providing stream processing are provided. To facilitate efficient query execution, the model index provides an interface for describing the hierarchical relationship between models to the query language. Based on query statistics and model performance, an alternative configuration of models might be proposed by a model advisor [71]. The advisor selects a configuration of models to use by utilizing a set of heuristics named indicators. Each indicator is either focused on a single time series or the relationship between multiple time series in the dimensional hierarchy. The choice of one of multiple models is determined based on a trade-off between accuracy and performance cost. The produced model configuration can then be loaded into $F^2$DB.

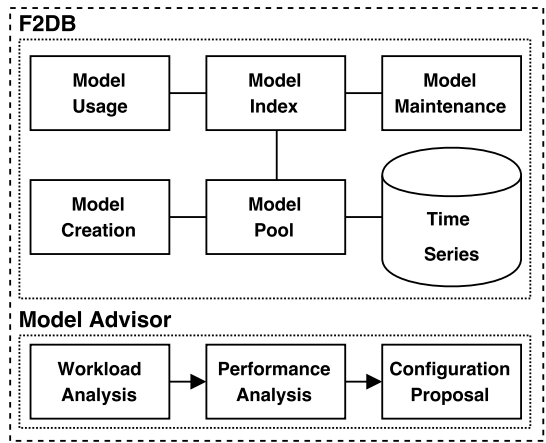

Fig. 14. Overview of the $F^2$DB TSMS. Models are constructed from time series stored in the base tables, indexed, and then used by the model usage component for query processing. The figure was redrawn from [70]

**Bakkalian et al.** [72] present a proof-of-concept PL/SQL extension for the Oracle RDBMS. The implemented extension allows time series to be stored and queried as linear functions in an OLAP data model. The system splits storage of a time series into two tables: the raw data is organized in one table as events with each row representing an event with a timestamp, metadata, and a value. Another table stores the intervals between each two consecutive events as a model, with the current implementation supporting linear functions only. Queries are executed directly against the models as they substitute the raw data representation. The use of linear functions allows the system to interpolate values not present in the raw data set to support AQP. The authors also discuss the theoretical use of models for forecasting time series as a benefit of a model-based approach to time series storage and querying. However, no information is provided about support for forecasting being implemented. In addition, no support for stream processing is described in the paper. The proposed system builds directly upon the following two publications in which some of the authors participated. Bebel et al. [73] proposed a model for OLAP analysis of sequential data that formalized the notion of sequences as facts and where sequences were formalized as an ordered collection of events. Building upon the model proposed in the previously mentioned paper, Koncilia et al. [74] proposed a new model enabling OLAP analysis of sequential data by formally introducing the notion of intervals as the time between two consecutive events, as well as defining a sequence to be an ordered set of intervals.

### 5.3 Discussion

TimeTravel [68] and $F^2$DB [70] provide similar capabilities for forecasting time series stored in an RDBMS. However, the intended use case for the systems differs. TimeTravel is focused on approximating time series using models to provide AQP for historical, present, and future values. In addition, a hierarchy of models is created so the system can select a model based on the necessary accuracy to decrease query time, and allow users to indicate what seasonality the time series will exhibit. $F^2$DB is motivated by a need to perform forecasting in a data warehouse and use the multi-dimensional hierarchy as part of the modelling process. The systems also differ in-terms of their ability to perform AQP as TimeTravel supports execution of AQP queries on both historical data and as forecast queries, while $F^2$DB's focus is on forecast queries. The system by Bakkalian et al. [72] represents time series using models but does not focus on their use for forecasting. This system instead uses models to reduce the storage requirement and query response time for OLAP queries. In addition to extending PostgreSQL and Oracle RDBMS with new methods for storing and querying time series, researchers have also extended RDBMSs with functionality for analytics without any changes to data storage through for example MADLib [83] and SearchTS [84]. Despite all of surveyed TSMSs in this section extend an RDBMS with AQP functionality, none of the systems implement functionality for stream processing.

## 6 FUTURE RESEARCH DIRECTIONS

The increase in time series data produced has lead researchers to discuss new directions for research into development of TSMSs, with new research opportunities being created as a result of this effort.

### 6.1 Research Directions Proposed in Literature

**Cuzzocrea et al.** [85] detail how the combination of sensor networks with cloud infrastructure enables the development of new types of applications, detail problems with the current state-of-the-art, and present a set of open problems regarding the combination of sensor networks with cloud infrastructure. The first problem presented is that different sources of sensor data often are heterogeneous, increasing the difficulty of combining different data sources for analytics. Heterogeneous data sources also complicate the addition of structure to a data set as a prerequisite for an analytical task. Another problem described is the amounts of sensor data produced which add additional requirements to the scalability of TSMSs. The authors propose to utilize methods from both RDBMSs and NoSQL database systems, and develop new systems with a high focus on scalability, primarily through horizontal partitioning and by reducing the reliance on join operations.

**Cuzzocrea** [86] provides an overview of the state-of-the-art methods for managing temporal data in general. However, while the general topic of the paper is temporal data management, a discussion of existing research and future research direction in the area of sensor data and scientific data management is also included in the paper. In addition, the author provides a discussion of additional open research questions within the area of temporal and sensor data management. For this summary only the proposed research directions relevant for management of time series data are included. The presented research directions are primarily centered around the problem of scalability. First the author argues that new storage methods are required, in conjunction with additional research into indexing mechanisms. In terms of data analysis, the author sees a hybrid system combining in-memory and disk based storage together with the next generation of system using the distributed data processing model popularized by Apache Hadoop as an import research direction. This architecture was later realized for general purpose data by SnappyData [87], [88]. In addition, the author proposes that further research into AQP method be performed, as such methods have already been proven successful for sensor data. The development of innovative analytical methods robust against imprecise values and outliers is presented as a pressing concern, and new methods for data visualization are presented as being necessary, as existing solutions for visualization of time series at different resolutions cannot be used at the scale required today.

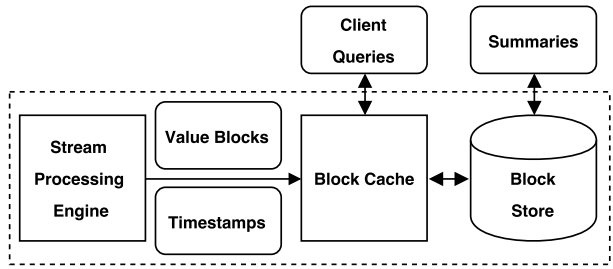

Fig. 15. Architecture proposed by Shafer et al. [12], redrawn from [12]

Additional arguments for further research into the development of TSMSs were presented by **Shafer et al.** [12]. Traditional RDBMSs support functionality unnecessary for processing time series, such as the ability to delete and update arbitrary values, with the argument that supporting functionality that is not necessary for analyzing time series adds complexity and overhead to these systems. In addition, since time series analysis often follows a common set of access patterns, creating a system optimized specifically for these access patterns should lead to a performance increase over a general purpose DBMS. The authors present an overview of some existing TSMSs and DBMSs suitable for time series analysis, and argue that they all effectively function as column stores with additional functionality specific for processing time series added on top. Using a column store for storing and processing time series provides the benefit of run length compression and efficient retrieval of a specific dimension of a time series [89]. However, the authors dispute the use of column stores and describe a set of design principles for a time series data store as shown in Fig. 15. They propose that time series data store should separate time series into timestamps and values, the values be partitioned based on their origin to make appending them to a time ordered list trivial, and last the values should be archived in blocks ordered by time to allow for efficient access to a continuous part of the time series. A preliminary evaluation demonstrates that using these principles, a higher compression rate compared to existing RDBMS and TSMS, can be achieved.

**Sharma et al.** [1] present new research directions for time series analysis in the area of Cyber-Physical Systems (CPSs). Creating CPSs by combining sensor networks with software for detailed analysis and decision-making allows for a new wave of informed automation system based on sensor readings. However, the authors argue that before such systems can be realized at a large scale, multiple challenges must be resolved. First, the lack of information about how changes to a CPS change the sensor readings produced, prevents systems from automatically controlling the monitored system based on information from the sensors. As a possible solution the authors propose using graph-based methods to infer the relationship between changes to a system and its sensor readings. Changes in discrete state could also be used to predicate changes to sensor readings, such as predicting a change in the speed of a car if cruise control is turned on or off. Another problem presented is scaling systems large enough to analyze the amount of low quality data large networks of cheap sensors produce, due both to the amount of data produced and the necessity to clean it before it can be analyzed. For this problem, the authors propose that existing methods designed for detection of sequential patterns in symbolic data could be effective for time series data mining, if efficient methods for converting time series to a symbolic representation can be created.

Another proponent for the development of specialized systems for analysis of time series and data series is **Palpanas** [3], [4], [5] as he argues that the development of a new class of systems he names Data Series Management System (SMS) is a requirement for analysis of data series to continue to be feasible as the amounts of data from data series increase. However, for an SMS to be realized, methods enabling a system to intelligently select appropriate storage and query execution strategies based on the structure of the data is necessary. As time series from different domains exhibit different characteristics, the system should be able to take these characteristics into account, and utilize one or multiple different representations to minimize the necessary storage requirement while still allowing for adaptive indexes to be created and queries to be executed efficiently. Parallel and distributed query execution must be an integrated part of the system to take advantage of not only multi-core processor and SIMD instructions, but also the scalability provided by distributed computing. To determine an appropriate execution method for a query in an SMS a cost-based optimizer specialized for data series would also be a required component, and it should abstract away the parallel execution of the analysis and indexing methods implemented in the system. However, the author argues that no existing cost-based optimizer is suitable for use in a SMS. In addition to the proposed system architecture, an argument for the development of a standardized benchmark for comparison of SMSs is also presented.

## 6.2 Online Model-Based Sensor Warehouse

In addition to the proposed research direction into storage and analysis of time series, and based on the surveyed systems, we see the following areas as being of primary interest for the scalability and usability of TSMSs to be further increased. The end result of the research we propose is a distributed TSMS with a physical layer storing time series as mathematical models, serving as the DBMS for a distributed data warehouse designed specifically for analysis of sensor data collected as time series and updated in real-time.

**Online Distributed Model-Based AQP:** Among the systems in the survey, few utilized models to approximate time series and provide a means for AQP, opting instead to reduce query response time through pre-computed aggregates and distributed computing. However, we believe that for use cases where approximate results are acceptable, the use of models provide a more flexible means for AQP. This is due to the configurable error bound and storage requirements, the capability to infer missing values and remove outliers, and due to some queries being answerable directly from the models, dramatically decreasing query response time. Similarly, to the use of models for distributed batch processing, the use of models for distributed stream processing should be evaluated, with the possibility that new algorithms could be developed due to the reduction in bandwidth used. Research in this direction has been performed by Guo et al. [51], [52], [53]. However, the proposed system does primarily demonstrate a new indexing solution. In addition, the system only provides limited support for low latency stream processing and the system does not implement a declarative query language. Centralized TSMSs that use models as an integral part of the system and provide a declarative interface have been designed, for example Plato [38], TimeTravel [68] and F$^2$DB [70]. However, as these systems do not utilize distributed computing, they are limited in terms of scalability.
**Automatic Model Management:** Additional support for helping users determine how model-based AQP should be utilized is necessary. Most TSMSs using model-based AQP either provide no means for implementing and selecting models appropriate for a particular time series or relegate the task to domain experts [38], with only a few systems, for example F$^2$DB [70], [71], providing a mechanism for automatic selection of models. While development of new models and integration of such into a TSMS properly will need to be performed manually for some time, a TSMS supporting AQP using models should at minimum implement methods for automatically fitting models to time series based on an error function for both batch and streaming data. While researchers have proposed some methods for automatically fitting models to time series stored in a DBMS, only a few publications present methods for approximating times series by automatically selecting appropriate models in real-time [90], [91]. Inspired by similar proposals in the area of machine learning [92], we propose that methods and systems be developed that interactively guide non-expert user to select appropriate models for a particular time series without the need for an exhaustive search when data is stored in a DBMS, and provide efficient selection of models for segmenting and approximating time series with a pre-specified error bound when ingested. Similarly, such a TSMS should as part of query processing provide insights into how an error bound should be interpreted and how a model might have effected the time series it represents.
**Model-Based Sensor Data Warehousing:** A distributed TSMS using models for stream and batch processing of time series has the capability to scale through the use of distributed processing by adjusting the acceptable error bound for a query, through selection of alternative models, and by implementing a query optimizer capable of taking the properties of the models used into account. However, to enable simple analytics on top of the model, an abstraction is needed. One possible abstraction, the data cube, has already been used successfully for analyzing time series data with the MIRABEL project as an example [69], [82], while the methods proposed by Perera et al. [77] and Bakkalian et al. [72] demonstrated that, for centralized systems, representing time series using models in a data cube can lead to reduced storage requirements and query response times. However, a system utilizing models for representing time series in a data cube has, to the best of our knowledge, never been successfully integrated with methods for continuously updating a data cube such as for example Stream Cubes [93] or systems such as Tidalrace [94], in a distributed system. Another benefit of using a data cube is that it is a well-known by data analysts, providing users of our envisioned TSMS with a familiar interface for analytics of multi-dimensional time series. Therefore, we see a data cube as a useful abstraction for analyzing multi-dimensional time series represented using models at the physical layer, if methods for continuously maintaining an OLAP cube can be unified with the methods for representing time series using models in a distributed TSMS.

# 7 CONCLUSION

The increasing amount of time series data that is being produced requires that new specialized TSMSs be developed. In this paper we have provided a systematic and thorough analysis and classification of TSMSs. In addition, we have discussed our vision for the next generation of systems for storage and analysis of time series, a vision that is based on the analysis in this survey and the directions for future work proposed by other researches in the field.

From our analysis we provide the following conclusions. TSMSs that use an *internal data store* and integrate it directly with the query processing component are predominately centralized systems, while distributed TSMSs are being developed using existing DFSs and distributed DBMSs. Research into systems with an internal data store is instead primarily focused on systems for embedded devices or which function as a proof-of-concept implementation of a new theoretical method. This point is further reinforced as only a few of the proposed TSMSs in the internal data store category can be considered mature enough for deployment in a business critical scenario. This situation is contrasted by the set of systems using *external data stores*, as a larger portion of these systems are distributed and developed by or in collaboration with a corporate entity where the system then can be deployed in order to solve a data management problem. As none of the TSMSs surveyed use a distributed storage system developed and optimized exclusively for persistent storage of time series, it is an open question what benefit such a storage solution would provide. Last, systems built as *extensions to existing RDBMSs* are few and all extend a RDBMS with support for AQP through the use of mathematical models. Other extensions to RDBMSs have added new methods for query processing but no additional functionality for storage of time series. The described RDBMSs are in general complete in terms of development, two of the systems were integrated into larger applications for use in a smart grid and the last being a prototype. However none of the RDBMS systems are distributed, limiting their ability to scale.

Additional functionality built on top of the central storage and query processing components is scarce in all three categories of systems. At the time of writing, only a limited number of systems provide any capabilities for stream processing to support transformation of the data as it is being ingested. In contrast to implementing stream processing during data ingestion, some TSMSs provide mechanisms for piping the data through a set of transformations after it has been stored in the TSMS or structure APIs around streams. Compared to stream processing, more TSMSs implement methods for approximating time series as a means to reduce query response time, storage requirements, or provide advanced analytical capabilities. However, for the systems that implement methods for approximating time series, it is uncommon to have an interface for domain experts to extend the systems with additional user-defined methods or models optimized for a particular domain or data set.

The future research directions proposed by experts in the field are primarily focused on the need for a TSMS with a storage solution and query processor developed from the ground up for time series, instead of reusing existing components or data models from for example an RDBMS for storage of time series. Additionally, it is proposed that such systems should support in-memory, parallel and distributed processing of time series as a means to reduce query processing time enough for interactive data analytics and visualization. AQP is mentioned as another possibility for reducing query processing time. As future work we propose that a TSMS that inherently provides a data cube as the means for analyzing multi-dimensional time series be developed. The system should provide interactive query speeds through the use of distributed in-memory processing and incorporate AQP as a central part of the system from the start. The system must support stream processing to transform and clean the data as it is ingested, and provide the means for construction of user-defined models, in order to compress the time series and enable AQP.

In summary, we propose that a distributed TSMS providing the same analytical capabilities as a data warehouse be developed specifically for use with time series. The TSMS should provide functionality that allows the data to be updated in real-time, support stream processing using user-defined functions and allow queries to be executed on both historical and incoming data at interactive query speed through the use of AQP.

## ACKNOWLEDGMENTS

This research was supported by the DiCyPS center funded by Innovation Fund Denmark [95].

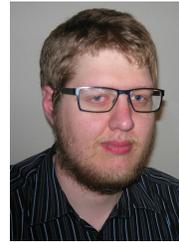

**Søren Kejser Jensen** is a PhD Student working on large scale sensor data management in real-time at the Center for Data-Intensive Systems (Daisy) at the Department of Computer Science, Aalborg University, Denmark. His research interests span multiple areas of Computer Science, including programming language design, compiler design, parallel and distributed programming, big data, data warehousing and extract-transform-load processes.

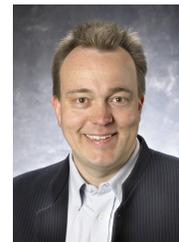

**Torben Bach Pedersen** is a professor at the Center for Data-Intensive Systems (Daisy) at the Department of Computer Science, Aalborg University, Denmark. His research concerns business intelligence and big data, especially "Big Multidimensional Data"–The integration and analysis of large amounts of complex and highly dynamic multidimensional data. He is an ACM Distinguished scientist, a senior member of the IEEE, and a member of the Danish Academy of Technical Sciences.

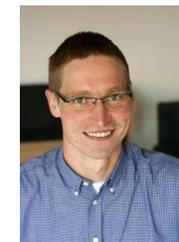

**Christian Thomsen** is an associate professor at the Center for Data-Intensive Systems (Daisy) at the Department of Computer Science, Aalborg University, Denmark. His research interests include big data, business intelligence, data warehousing, and extract-transform-load processes.